\documentclass{elsart}

\input epsf

\usepackage{amssymb}

\begin{document}

\begin{frontmatter}



\title{Lattice animals on a staircase and generalized Fibonacci
numbers}


\author{Lo\"\i c Turban}

\address{Laboratoire de Physique des Mat\'eriaux, UMR CNRS 7556,\\ 
Universit\'e Henri Poincar\'e, BP~239,\\  
F--54506~Vand\oe uvre l\`es Nancy Cedex, France\\
e-mail: \tt turban@lpm.u-nancy.fr}

\begin{abstract}
We study the statistics of column-convex lattice animals generated by the
stacking of squares on a staircase with step height $p$. We calculate the
number of animals with area $k$ living on $l$ stairs. The total number of
animals with area $k$ is given by the generalized Fibonacci number
$F_p(k)$. Exact results for the mean length and mean height of animals with area
$k$ are also obtained and we examine their asymptotic behaviour.   
\end{abstract}

\end{frontmatter}
\section{Introduction}
\label{}
A  lattice animal is a connected cluster of occupied
sites on a lattice or the corresponding cluster of occupied cells on the dual
lattice, also called a polyomino. In two
dimensions, animals can be enumerated either according to their area,  
the number of sites belonging to a cluster, or according to their
perimeter, the number of vacant sites which are first neighbours of
occupied sites. 

The enumeration problem is quite difficult in general. Only some
bounds on the asymptotic behaviour are known \cite{klarner73}. For restricted
classes of animals such as Ferrers graphs, convex and/or directed animals,
some exact results have been obtained during the last two
decades~\cite{bousquet96a,brak90,dhar82a,dhar82b,hakim83,joyce94,lin91,nadal82}.
A review of the state of the problem before 1996 can
be found in~\cite{bousquet96b}. 

In the present work, we consider lattice animals resulting from the stacking
of squares on a staircase with step height $p$. Two square faces are
connected when they share an edge as shown in figure 1. Such
animals are called column-convex or vertically convex because the 
intersection of a vertical line with the perimeter generates at most two
connected components. The enumeration problem and the morphology of the animals
involve the generalized Fibonacci numbers $F_p(k)$. The case $p=0$, 
which corresponds to the stacking of squares on a line, is well known
(see~\cite{privman89} for example). With $p=1$, a case recently considered
in~\cite{turban2000}, one obtains a relation with the ordinary Fibonacci numbers
$F_k=F_1(k)$. A similar connection has been noticed a long time ago between 
column-convex
directed animals and ordinary Fibonacci numbers with odd
indices~\cite{klarner65}

In section 2, the number of animals with area $k$ living on $l$ stairs is
deduced from their generating function. Their total number is given by the
generalized Fibonacci numbers $F_p(k)$. In sections~3 and~4, we obtain exact
expressions for the mean length and mean height of an animal with area $k$.
The asymptotic behaviour is discussed in section~5.   


\begin{figure}[t]
\epsfxsize=.8\textwidth
\begin{center}
\mbox{\epsfbox{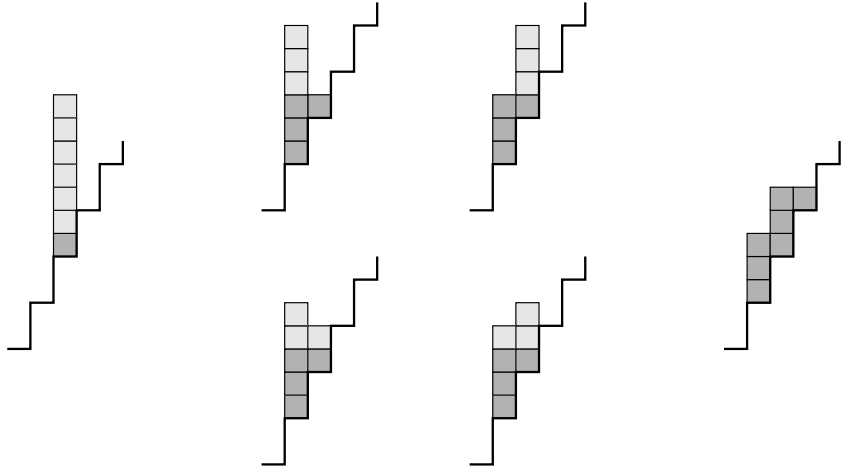}}
\end{center}
\caption[]{The 6 different animals obtained by stacking
7 squares on $l=1,2,3$ stairs in a staircase with step height 2. The $3l-2$
dark squares build the minimal cluster necessary to cover $l$ stairs.} 
\label{fig-1}
\end{figure}

\section{Number of animals}
We consider animals with area $k$ living on $l$ stairs in a staircase
with step height $p$ as shown in figure~1. In order to calculate their number
$F_p(k,l)$, we introduce the generating function     
\begin{equation}
F_p(z,t)=\sum_{k,l=1}^\infty F_p(k,l)\, z^kt^l
\label{e2.1}
\end{equation}
which satisfies the following relation:
\begin{eqnarray}
F_p(z,t)&=&t\,(z+z^2+\cdots+z^p)+t\,\frac{z^{p+1}}{1-z}\,[1+F_p(z,t)]\nonumber\\
&=&tz\,\frac{1-z^p}{1-z}+t\,\frac{z^{p+1}}{1-z}\,[1+F_p(z,t)]\,.
\label{e2.2}
\end{eqnarray}
The first term corresponds to animals with up to $p$ squares stacked on a
single stair. In the second term, the first factor corresponds to a column
with more than $p$ squares on the lowest stair, such that the animal may or not
continue to grow on the next stair, as indicated in the second factor.

Equation~(\ref{e2.2}) leads to the generating function
\begin{equation}
F_p(z,t)=\frac{tz}{1-z-tz^{p+1}}
\label{e2.3}
\end{equation}
which can be expanded to give:
\begin{eqnarray}
F_p(z,t)&=&tz\sum_{m=0}^\infty z^m(1+tz^p)^m
=\sum_{m=0}^\infty\sum_n{m\choose n}z^{m+pn+1}t^{n+1}\nonumber\\
&=&\sum_{l\geq1}\sum_{k=(p+1)(l-1)+1}^\infty{k-p(l-1)-1\choose l-1}\, z^kt^l\,.
\label{e2.4}
\end{eqnarray}

Comparing equations~(\ref{e2.1}) and~(\ref{e2.4}), the number
of animals with area $k$ living on $l$ stairs is given by
\begin{equation}
F_p(k,l)={k-p(l-1)-1\choose l-1}
\label{e2.5}
\end{equation}
when $k\geq (p+1)(l-1)+1$ and vanishes otherwise.
This expression can be rewritten as
\begin{equation}
F_p(k,l)={l+m-1\choose l-1}\,,\quad m=k-k_{min}=k-[(l-1)(p+1)+1]\,,
\label{e2.6}
\end{equation}
where $k_{min}$ is the number of squares needed to cover
$l$ successive stairs with a minimal connected cluster forming the basis of an
animal, as shown in figure~1. Thus $F_p(k,l)$ is also the number of
unconstrained configurations of the remaining $m=k-k_{min}$ squares, placed on
top of this basis, which is given by the binomial coefficient in~(\ref{e2.6}). 

{}From the addition/induction relation for the binomial
coefficients, 
\begin{displaymath}
{m+1\choose n+1}={m\choose n+1}+{m\choose n}
\label{e2.7}
\end{displaymath}
we deduce the following recursion relation:
\begin{equation}
F_p(k+1,l+1)=F_p(k,l+1)+F_p(k-p,l)\,,\qquad k\geq pl+1\,.
\label{e2.8}
\end{equation}

The total number of animals with area $k$ is given by:
\begin{equation}
F_p(k)=\sum_{l=1}^\infty F_p(k,l)=\sum_{0\leq m\leq{k-1\over
p+1}}{k-pm-1\choose m}\,.
\label{e2.9}
\end{equation}
Its generating function is obtained by setting $t=1$ in~(\ref{e2.3}) and
reads: 
\begin{equation}
F_p(z)=\frac{z}{1-z-z^{p+1}}\,.
\label{e2.10}
\end{equation}
According to~(\ref{e2.8}) and~(\ref{e2.9}), $F_p(k)$ satistifies the
recursion relation for the generalized Fibonacci
numbers:
\begin{displaymath}
F_p(k+1)=F_p(k)+F_p(k-p)\,,\quad F_p(1)=\cdots=F_p(p+1)=1\,.
\label{e2.11}
\end{displaymath}
\section{Mean length}
We define the mean length of an animal with area $k$ as the
mean number of successive stairs it occupies:
\begin{equation}
\overline{l_p}(k)=\frac{A_p(k)}{F_p(k)}\,,\qquad
A_p(k)=\sum_{l=1}^\infty l F_p(k,l)\,.
\label{e3.1}
\end{equation}
Introducing the generating function
\begin{displaymath}
A_p(z)=\sum_{k=1}^\infty A_p(k)z^k
\label{e3.2}
\end{displaymath}
and making use of equations~(\ref{e2.1}), (\ref{e2.3}) and~(\ref{e2.10}), we
obtain: 
\begin{eqnarray*}
A_p(z)&=&\left.\frac{\partial F_p(z,t)}{\partial t}\right\vert_{t=1}
=\frac{z}{1-z-z^{p+1}}+\frac{z^{p+2}}{(1-z-z^{p+1})^2}\nonumber\\
&=&\frac{1-z}{z}\left(\frac{z}{1-z-z^{p+1}}\right)^2=(z^{-1}-1)\, F_p^2(z)\,.
\label{e3.3}
\end{eqnarray*}
With $F_p^2(z)=\sum_{k=2}^\infty F_p^{(2)}(k)z^k$, the last equation leads to: 
\begin{equation}
A_p(k)=[z^k]\,(z^{-1}-1)\, F_p^2(z)
=F_p^{(2)}(k+1)-F_p^{(2)}(k)
\label{e3.4}
\end{equation}
where $[z^k]\,f(z)$ is the coefficient of $z^k$ in
$f(z)$. According to~(\ref{e2.10}) we have
\begin{displaymath}
(1-z-z^{p+1})\,F_p^2(z)=\frac{z^2}{1-z-z^{p+1}}=zF_p(z)
\label{e3.5}
\end{displaymath}
so that:
\begin{eqnarray}
F_p(k)&=&[z^k]\,F_p(z)=[z^k]\,(z^{-1}-1-z^p)\,F_p^2(z)\nonumber\\
&=&F_p^{(2)}(k+1)-F_p^{(2)}(k)-F_p^{(2)}(k-p)\,.
\label{e3.6}
\end{eqnarray}
Hence, combining (\ref{e3.1}), (\ref{e3.4}) and~(\ref{e3.6}), the mean
length can be rewritten as:
\begin{equation}
\overline{l_p}(k)=1+\frac{F_p^{(2)}(k-p)}{F_p(k)}\,.
\label{e3.7}
\end{equation}

$F_p^{(2)}(k)$ can be expressed in terms of $F_p(k)$ using the
relation 
\begin{displaymath}
z^2\,\frac{\d}{\d z}\left[\frac{F_p(z)}{z}\right]=[1+(p+1)\,z^p]\,
F_p^2(z)
\label{e3.8} 
\end{displaymath}
which follows from~(\ref{e2.10}) and leads to:
\begin{displaymath}
(k-1)\,F_p(k)=F_p^{(2)}(k)+(p+1)\,F_p^{(2)}(k-p)\,.
\label{e3.9}
\end{displaymath}
Replacing $F_p^{(2)}(k-p)$ by its value taken from~(\ref{e3.6}), one
obtains:
\begin{displaymath}
(p+1)\,F_p^{(2)}(k+1)-p\,F_p^{(2)}(k)=(k+p)\,F_p(k)\,.
\label{e3.10}
\end{displaymath}
Thus we have the recursion relation
\begin{displaymath}
F_p^{(2)}(k)=\frac{k+p-1}{p+1}\,F_p(k-1)+\frac{p}{p+1}\,F_p^{(2)}(k-1)\,,\quad
F_p^{(2)}(1)=0\,,
\label{e3.11}
\end{displaymath}
which can be iterated to give:
\begin{eqnarray}
F_p^{(2)}(k)&=&\sum_{n=1}^{k-1}\left(\frac{p}{p+1}\right)^n\,\frac{k+p-n}{p}
\,F_p(k-n)\nonumber\\
&=&\left(\frac{p}{p+1}\right)^k\,\sum_{n=1}^{k-1}\left(\frac{p+1}{p}
\right)^n\,\frac{p+n}{p} \,F_p(n)\,,\qquad k\geq2\,. 
\label{e3.12}
\end{eqnarray}
Together with equation~(\ref{e3.7}), this leads to the mean length in terms of
generalized Fibonacci numbers.

When $p=1$ the coefficients $F_1^{(2)}(k)$ can be obtained more
directly than above for a general value of $p$ 
(see reference~\cite{graham94}, p. 353). They are given by
\begin{displaymath}
F_1^{(2)}(k)=\frac{2k\,F_1(k+1)-(k+1)\,F_1(k)}{5}
\label{e3.13}
\end{displaymath}
which, together with~(\ref{e3.12}) for $p=1$, leads to the following
identity for ordinary Fibonacci numbers: 
\begin{displaymath}
\sum_{n=1}^{k-1}2^n(n+1)\,F_1(n)=2^k\,\frac{2k\,F_1(k+1)-(k+1)\,F_1(k)}{5}\,.
\label{e3.14}
\end{displaymath} 

\section{Mean height}
The mean height of an animal with area $k$, measured from the staircase, is
defined as:
\begin{displaymath}
\overline{h_p}(k)=\frac{B_p(k)}{F_p(k)}\,,\quad B_p(k)=\sum_{l=1}^\infty
\frac{k}{l}\,F_p(k,l)=\sum_{l=1}^\infty\frac{k}{l}{k-p(l-1)-1\choose l-1}\,.
\label{e4.1}
\end{displaymath}
The absorption/extraction identity gives:
\begin{displaymath}
{k-p(l-1)\choose l}=\frac{k-p(l-1)}{l}{k-p(l-1)-1\choose l-1}\,,\quad l\geq1\,.
\label{e4.2}
\end{displaymath}
Hence we have
\begin{equation}
B_p(k)=\frac{kp}{k+p}\sum_l{k-p(l-1)-1\choose l-1}
+\frac{k}{k+p}\sum_{l\geq1}{k-p(l-1)\choose l}
\label{e4.3}
\end{equation}
where the last sum can be rewritten as:
\begin{eqnarray}
\sum_{l\geq1}{k-p(l-1)\choose l}&=&\sum_{l}{k-p(l-1)\choose l}-1\nonumber\\
&=&\sum_{l}{k+p+1-p(l-1)-1\choose l-1}-1\,.
\label{e4.4}
\end{eqnarray}
Taking into account the combinatorial definition of the generalized Fibonacci
numbers in~(\ref{e2.9}), equations~(\ref{e4.3}) and~(\ref{e4.4}) yield
\begin{displaymath}
B_p(k)=\frac{k}{k+p}\,[F_p(k+p+1)+p\,F_p(k)-1]
\label{e4.5}
\end{displaymath}
so that the mean height is finally given by:
\begin{equation}
\overline{h_p}(k)=\frac{k}{k+p}\,\frac{F_p(k+p+1)+p\,F_p(k)-1}{F_p(k)}\,.
\label{e4.6}
\end{equation}

\section{Asymptotic behaviour}
In order to obtain the asymptotic behaviour of the
generalized Fibonacci numbers, the generating function $F_p(z)$ in
equation~(\ref{e2.10}) can be written as the partial fraction expansion
\begin{equation}
F_p(z)=\frac{C_p}{1-\phi_pz}+r(z)
=\sum_{k=1}^\infty C_p(\phi_pz)^k+r(z)
\label{e5.1}
\end{equation}
where $\phi_p$ is the reciprocal of the smallest root in modulus of the
denominator of $F_p(z)$, which is assumed to be a simple root. It satisfies the
relations:   
\begin{equation}
\phi_p^{p+1}=1+\phi_p^p=\frac{\phi_p}{\phi_p-1}\,.
\label{e5.2}
\end{equation}
In equation~(\ref{e5.1}), $r(z)$ gives the contribution of the $p-1$
other roots to the partial fraction expansion. It remains finite at
$z=\phi_p^{-1}$. At large $k$-values, the leading contribution to $F_p(k)$
comes from the first term so that  
\begin{equation}
F_p(k)\simeq C_p\phi_p^k\,,\qquad k\gg1\,.
\label{e5.3}
\end{equation}
The calculation of the amplitude $C_p$ proceeds as follows. Taking the derivative
of the inverse of $F_p(z)$, we have:
\begin{equation}
\left.\frac{\d F_p^{-1}(z)}{\d z}\right\vert_{z=\phi_p^{-1}}
=\lim_{z\to\phi_p^{-1}}\frac{F_p^{-1}(z)}{z-\phi_p^{-1}}=-\frac{\phi_p}{C_p}\,.
\label{e5.4}
\end{equation}
Thus, from~(\ref{e2.10}), we deduce
\begin{displaymath}
C_p=\frac{1}{1+(p+1)\,\phi_p^{-p}}=\frac{1}{(p+1)\phi_p-p}
\label{e5.5}
\end{displaymath}
where the last expression is obtained using~(\ref{e5.2}). It is easy to
verify that a multiple smallest root would lead to a vanishing derivative
in~(\ref{e5.4}), which justifies our previous assumption.

Let us now study the behaviour of the mean length at large $k$-values. For this
we need an asymptotic expression for $F_p^{(2)}(k)$. Equation~(\ref{e3.12}) leads
to:  
\begin{displaymath}
F_p^{(2)}(k)\simeq C_p\left(\frac{p}{p+1}\right)^k\,
\sum_{n=1}^{k-1}\frac{p+n}{p}\,\left(\frac{p+1}{p}\,\phi_p\right)^n\,.
\label{e5.6}
\end{displaymath}
Summing the series, to leading order, one obtains:
\begin{equation}
F_p^{(2)}(k)\simeq \frac{C_p}{(p+1)\phi_p-p}\, k\phi_p^k\,.
\label{e5.7}
\end{equation}
According to equations~(\ref{e3.7}), (\ref{e5.3}) and~(\ref{e5.7}), the mean length
behaves asymptotically as 
\begin{displaymath}
\overline{l_p}(k)=\frac{k}{p+\phi_p^{p+1}}+O(1)
=\frac{\phi_p-1}{(p+1)\phi_p-p}\,k+O(1)
\label{e5.8}
\end{displaymath}
where~(\ref{e5.2}) has been used to reduce the denominator.

For the mean height in equation~(\ref{e4.6}), we have the following asymptotic
form: 
\begin{displaymath}
\overline{h_p}(k)=p+\phi_p^{p+1}+O(k^{-1})
=p+\frac{\phi_p}{\phi_p-1}+O(k^{-1})\,.
\label{e5.9}
\end{displaymath} 

These results can be recovered and extended by considering the behaviour when
$k\gg1$ of $F_p(k,l)$ in equation~(\ref{e2.5}). It is obtained by expanding $\ln
F_p(k,l)$ to second order near its maximum, making use of the Stirling
approximation. The asymptotic distribution is Gaussian  and reads: 
\begin{displaymath}
F_p(k,l)\simeq \frac{C_p\phi_p^k}{\sqrt{2\pi\overline{\Delta l_p^2}}}
\exp\left[-\frac{(l-\overline{l_p}(k))^2}{2\overline{\Delta l_p^2}}\right]\,.
\label{e5.10}
\end{displaymath}
Here $\overline{l_p}(k)$ and 
\begin{displaymath}
\overline{\Delta l_p^2}=\frac{\phi_p(\phi_p-1)}{[(p+1)\phi_p-p]^3}\,k
\label{e5.11}
\end{displaymath}
are the leading contributions to the mean value of $l_p(k)$ and to its
mean-square deviation. The prefactor follows from the normalization to $F_p(k)$.

Finally let us examine some special values of $p$.

The case $p=0$ corresponds to the staking of squares on a flat
surface~\cite{privman89} for which
\begin{displaymath}
F_0(k)=2^{k-1}\,,\qquad
F_0^{(2)}(k)=(k-1)\,2^{k-2}\,,\qquad\phi_0=2\,. 
\label{e5.12}
\end{displaymath}
The mean length in~(\ref{e3.7}) and the mean height in~(\ref{e4.6}) behave
as:  
\begin{displaymath}
\overline{l_0}(k)=\frac{k+1}{2}=\frac{k}{2}+O(1)\,,\qquad
\overline{h_0}(k)=\frac{2^k-1}{2^{k-1}}=2+O(2^{-k})\,.
\label{e5.13}
\end{displaymath}

The statistics of animals on a staircase with $p=1$ is connected to the usual
Fibonacci numbers, $F_1(k)=1,1,2,3,5,8,\cdots$. Then, according to~(\ref{e5.2}),
$\phi_1=(1+\sqrt{5})/2$ is the golden mean $\phi$ and 
\begin{displaymath}
\overline{l_1}(k)=\frac{k}{1+\phi^2}+O(1)\,,\qquad
\overline{h_1}(k)=1+\phi^2+O(k^{-1})\,,
\label{e5.14}
\end{displaymath}
in agreement with the results of reference~\cite{turban2000}.

\end{document}